\documentclass[%
 reprint,
 amsmath,amssymb,
 aps,
]{revtex4-2}

\usepackage[table]{xcolor}
\usepackage{array}
\usepackage{graphicx}
\usepackage{dcolumn}
\usepackage{bm}

\begin{document}

\title{Apparent fractional charge signatures in PbTe quantum dots\\due to capacitively coupled charge trap dynamics}

\author{Seth Byard}
\author{Maksim Gomanko}
\author{Sergey M. Frolov}
\affiliation{
 Department of Physics and Astronomy, University of Pittsburgh, Pittsburgh, PA, 15260, USA
}

\date{\today}

\begin{abstract}
We report the observation of fractional shifts in the experimental stability diagrams of PbTe nanowire quantum dots. Although this behavior may appear to suggest fractional charge transport, akin to that reported in the fractional quantum Hall regime, the quasi-one-dimensionality of the system and absence of an applied magnetic field indicate that the presence of fractional charges is highly unlikely. We instead attribute these effects to the presence of one or more spurious dots, or charge traps, capacitively coupled to the primary dot. Our findings illustrate how signatures of fractional charge transport may be replicated through trivial mesoscopic Coulombic effects.
\end{abstract}

\maketitle

\section{Introduction}

Many important phenomena in quantum devices are related to, or revealed by, the motion of charges. This includes any and all electronic transport measurements which largely study free band electrons or holes. Behaviors of solid state qubits, even when interrogated by microwave radiation, ultimately reduce to charges changing position. Characterization and modeling of charge noise has been a persistent theme that has focused on $1/f$, telegraph, white or Johnson, as well as quantum noise~\cite{dutta, paladino}.

\begin{figure}[h]
\includegraphics[scale=0.345]{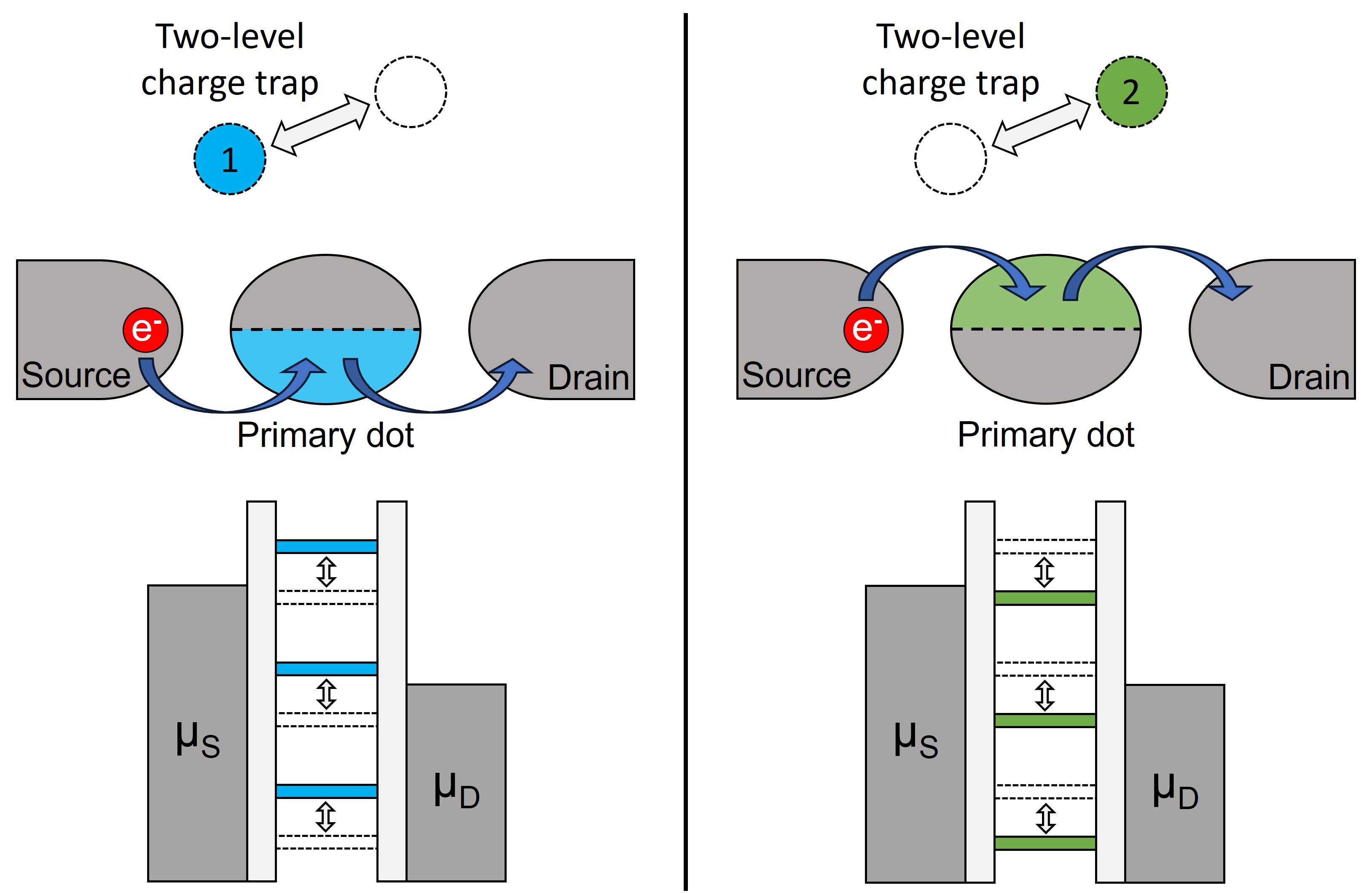}
\caption{\label{fig:dots_schematic} Schematic model for a primary quantum dot capacitively coupled to a charge trap, where a change in the trap's occupancy shifts electrochemical potential levels in the dot. The electron wavefunction inside the dot is shown here to be shifted between the upper and the lower part of the dot.}
\end{figure}

In nanostructures, carrier wavefunctions and their dynamics are sensitive to nearby charge distributions, whether those are externally imposed such as due to electrostatic gating, intrinsic such as dopant sites, or unintended such as charge traps and defects (Fig.~\ref{fig:dots_schematic}). When it comes to the unintended charge dynamics, gate voltage hysteresis and leakage has been linked to hopping of charges between the gate layer and the current-carrying layer~\cite{pioro-ladriere}. Charge traps changing occupation were studied as a source of dephasing in superconducting qubits, where they induce  critical current noise in Josephson tunnel junctions~\cite{vanHarlingten}. A related two-level system concept covers energy relaxation in microwave circuits~\cite{klimov, lisenfeld, burnett}. Charge traps are found within the bulk of various materials, such as defect or donor sites in semiconductors, within the oxide layers on the surfaces, and in molecules accidentally deposited on surfaces~\cite{zorin, culcer, connors}. 

Low-frequency charge jump events, such as the de-occupation and re-occupation of charge traps, can lead to visible shifts in experimental data when data acquisition time falls in-between the timescales for a charge to change position and the dwell time. This can be perceived as large and discrete jumps of the signal when the characteristic size of the system and the distance to a charge trap are comparable, i.e. when a trap is ‘nearby’ (Fig.~\ref{fig:dots_schematic}). This is common in measurements of semiconductor quantum dots, single-electron transistors, quantum point contacts, and nanowires~\cite{buehler, hitachi, zimmerli}.

On the other hand, measurements of larger samples, such as Hall bars, may not demonstrate charge jumps in the data, either because individual charge events average out, or in the quantum Hall regime, because the exact position of the edge state does not alter conductance. Electronic interferometers, however, are still sensitive to charge jumps because the interference of electron trajectories is strongly affected by the paths traveled, or by the enclosed total area~\cite{ji}.

Charge jumps are often perceived as obstructing the clear observation of the studied signals. Nevertheless, artificial removal of charge jump occurrences from the data post-acquisition does not have a solid justification~\cite{zhang}. Suppose a charge shift is removed from a gate trace of conductance: the position of the charge trap is generally not within the gate electrode, and the detailed electric field distribution imposed by the gate and the trap are not the same, hence the wavefunctions in the nanostructure are affected differently by the two. Artificially subtracting jumps due to charge traps from the signal can also create the false impression of a regime more stable in time than it actually is.

A distinct class of behaviors of high interest uses discrete signal changes due to charge jumps as a signature of novel and highly sought-after phenomena. One example is shifts in impedance detected in superconducting circuits, including in putative topological qubits, due to quasiparticle number or parity changes~\cite{microsoft}. Anyons known from the quantum Hall effect are characterized by fractional charges, such as $1/3$ of the elementary charge \cite{stern, stormer}. Experiments where abrupt shifts of approximately $1/3$ in periodic signals have been interpreted as providing direct evidence for such fractional charges~\cite{nakamura, werkmeister, samuelson}.

Here, we demonstrate that similar shifts in conductance patterns can be observed in regimes unlikely to host fractionalized quasiparticles. We study blockade oscillations in quantum dots defined by electrostatic gates in PbTe nanowires. Near zero magnetic field, we observe charge shifts of approximately $1/3$ the period. The shifts' positions are dependent on two gate electrodes and visually move diagonally through the conductance map, against the primary blockade oscillations. Our findings illustrate that charge shifts in periodic patterns can have trivial origins, and that caution should be exercised before considering non-trivial interpretations.

\section{Methods}

\begin{figure}[h]
\includegraphics[scale=0.295]{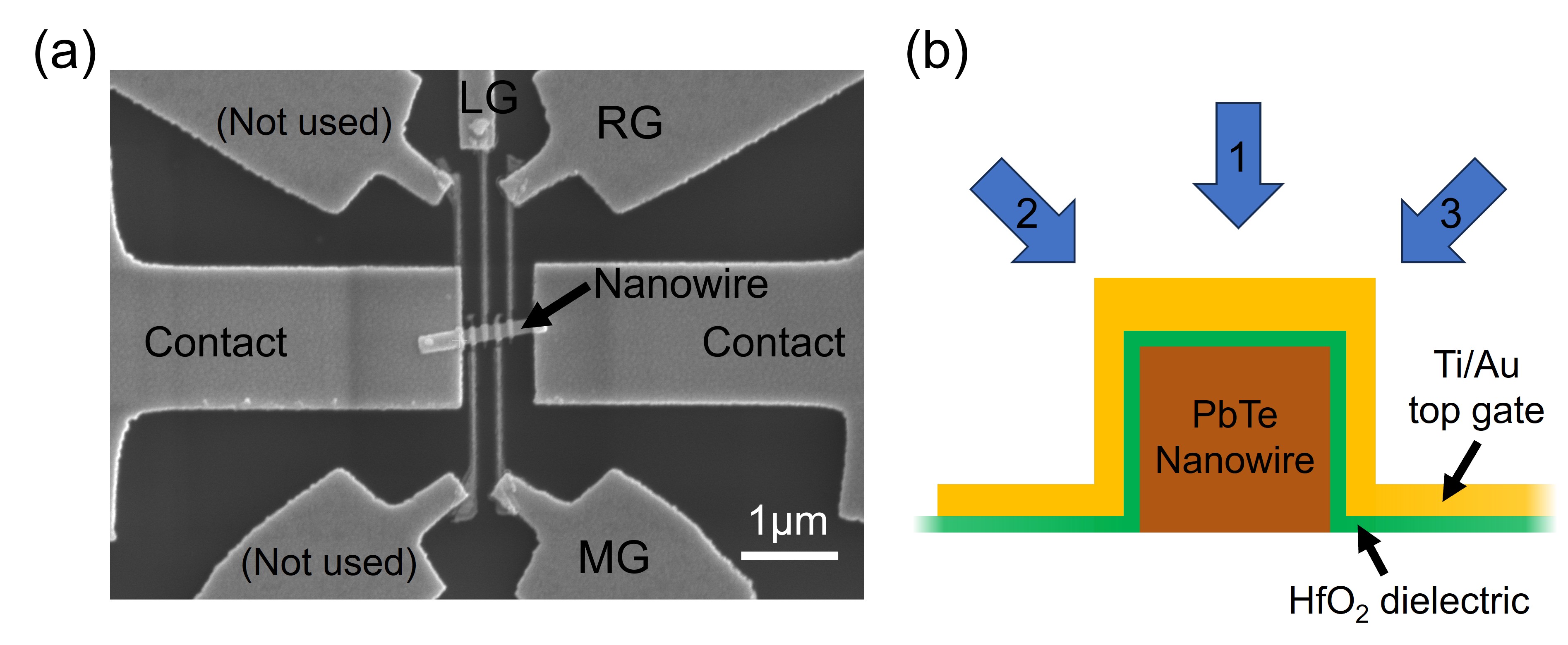}
\caption{\label{fig:device} Primary PbTe nanowire quantum dot device used in this study. (a) SEM image of the device indicating the nanowire, contacts, and the left, right and middle gates $LG$, $RG$ and $MG$. (b) Schematic diagram of the device's cross-section. Dielectric HfO\textsubscript{2} coats an exposed section of nanowire, and gates are then deposited on it at a series of angles.}
\end{figure}

The devices used in this study are fabricated with PbTe nanowires grown by molecular beam epitaxy (MBE) \cite{schellingerhout, gomanko}. The primary device is shown in Fig.~\ref{fig:device}. Standard electron beam lithography (EBL) and thin film deposition techniques are used to define 10/160 nm Ti/Au ohmic contacts. The samples are then uniformly coated in 10 nm HfO\textsubscript{2} deposited by atomic-layer deposition (ALD). Finally, a set of top gate electrodes is deposited at three different angles without breaking the vacuum. Although 5 gates were fabricated per device, only 3 gates were usable for the primary device of this study; these are labeled as "left," "middle," and "right" gates ($LG$, $RG$ and $MG$). 

Data are obtained from standard DC measurements done primarily in a dilution refrigerator with a base temperature of 100 mK. The measurements highlighted in this study were done in the absence of externally applied magnetic fields.

\section{Results}

\begin{figure}[h]
\includegraphics[scale=0.35]{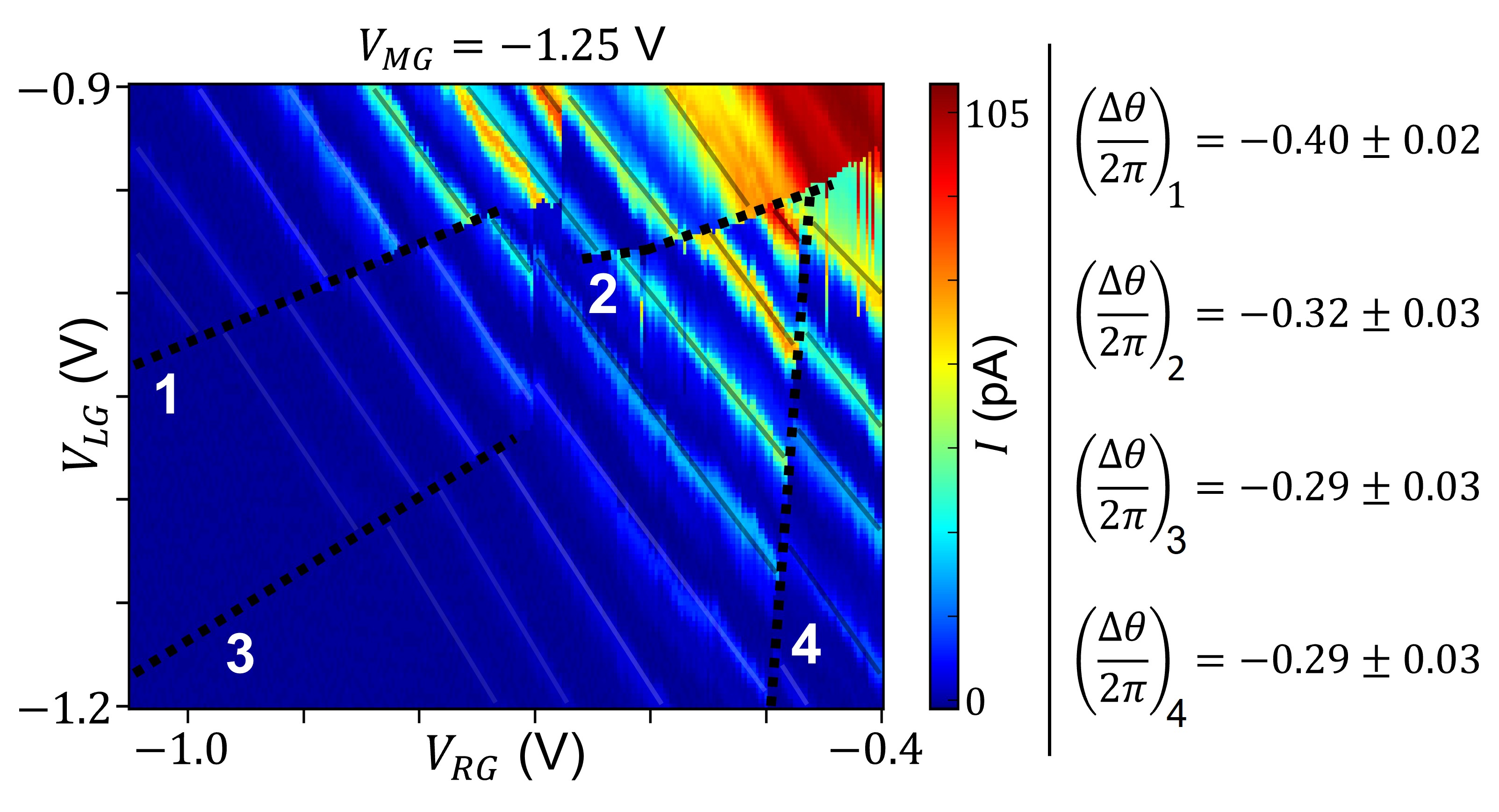}
\caption{\label{fig:data_main} Experimental quantum dot stability diagram as a function of left and right gate voltages, with the middle gate held at a fixed potential. Solid lines are added along the transport resonances as visual aids, and dashed black lines indicate select, abrupt shifts in this pattern. The magnitude of the $n^{th}$ shift, $(\Delta\theta/2\pi)_n$, is defined to be the average displacement of the resonances' midpoints along the direction of that shift, expressed as a fraction of the oscillation period. The magnitudes of the four indicated shifts are recorded to the right of the plot.}
\end{figure}

Fig.~\ref{fig:data_main} shows an experimental stability diagram of a PbTe nanowire quantum dot. The plot illustrates measured current as a function of left and right gate voltages, while a middle gate is held at a fixed potential. As the gates are swept, the dot oscillates between states of transport and blockade, or high and low conductance, respectively. Crossing through a high-conductance "stripe," or resonance, from one blockade state to another is equivalent to changing the dot's occupancy by one electron.

On top of the pattern of diagonal resonances, which is typical of a quantum dot, we observe apparent lines along which the pattern shifts by a fraction of its period. The magnitude of a shift is characterized by a value $\Delta\theta/2\pi$, where a shift by a full period would be equal to 1. We observe shifts close to the value of $1/3$ at several of these transitions. The shifts are neither vertical nor horizontal, as would be the case either for charge jumps related purely to one of the gates, or for another source of time-dependent noise unrelated to the device. Rather, they run diagonally against the primary oscillating pattern, indicating dependence on both gates. 

\begin{figure}[h]
\includegraphics[scale=0.34]{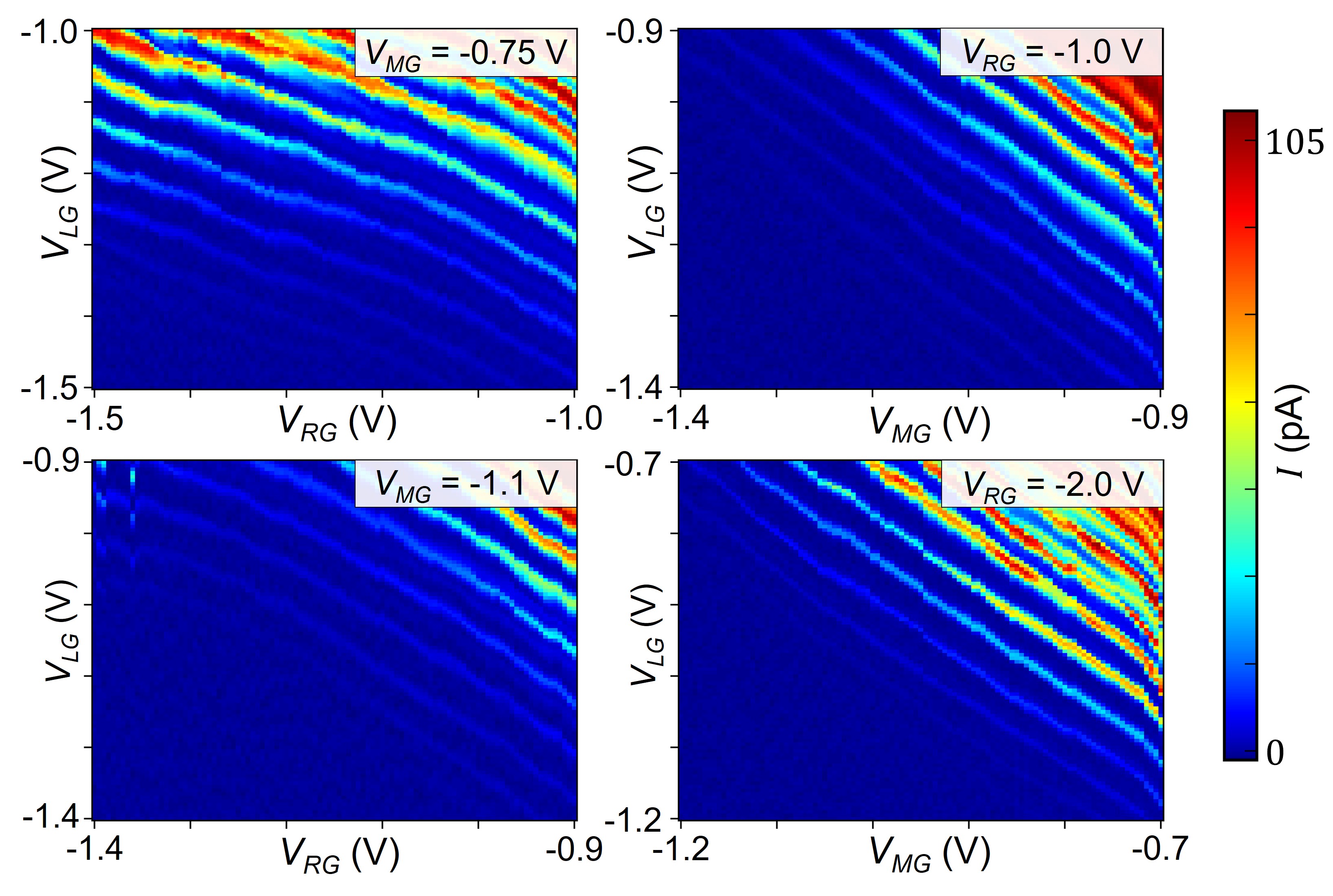}
\caption{\label{fig:data_boring} Typical stability diagrams which lack significant resonance shifts. Any observed shifting behavior is inconsistent and small in scale.}
\end{figure}

In contrast with results discussed thus far, Fig.~\ref{fig:data_boring} presents an assortment of experimental stability diagrams which lack any significant pattern shifts. These are obtained at diffrerent settings of the middle gate, but changes like this can generally be dependent on time and the history of sweeping gates.

\begin{figure}[h]
\includegraphics[scale=0.4]{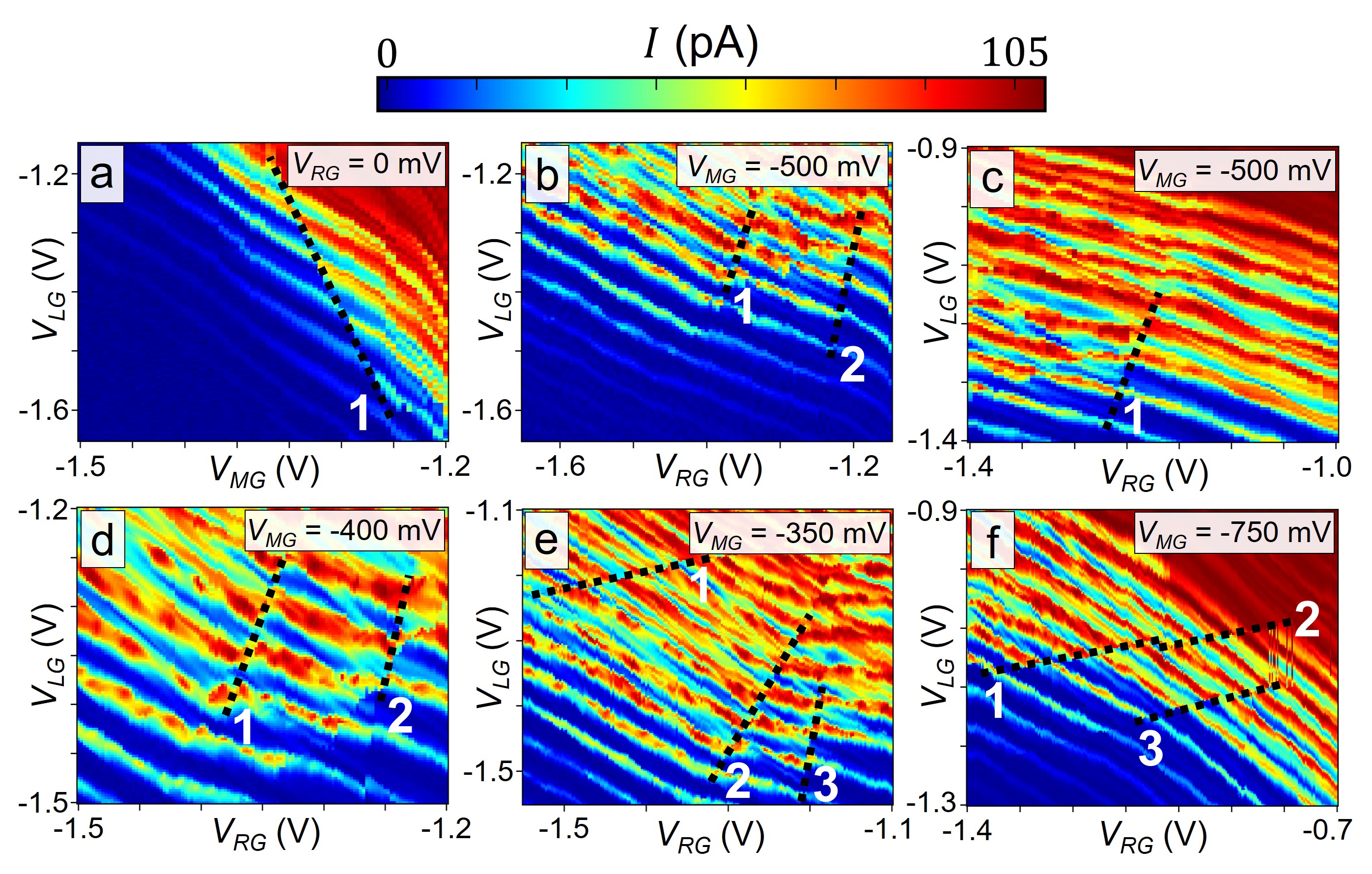}
\begin{center}
\setlength{\arrayrulewidth}{0.1mm}
\setlength{\tabcolsep}{5pt}
\renewcommand{\arraystretch}{1.2}
\begin{tabular}{|c|c|c|c|}
\hline 
\large{Subfigure} & \large{$\left(\frac{\Delta\theta}{2\pi}\right)_1$} & \large{$\left(\frac{\Delta\theta}{2\pi}\right)_2$} & \large{$\left(\frac{\Delta\theta}{2\pi}\right)_3$} \\[2pt]
\hline
a & $+0.21\pm 0.03$ & - & - \\
\hline
b & $-0.38\pm 0.05$ & $-0.27\pm 0.04$ & -\\
\hline
c & $-0.19\pm 0.03$ & - & - \\
\hline
d & $-0.38\pm 0.04$ & $-0.47\pm 0.07$ & -\\
\hline
e & $-0.61\pm 0.12$ & $-0.47\pm 0.07$ & $-0.30\pm 0.02$\\
\hline
f & $-0.40\pm 0.06$ & $-0.69\pm 0.02$ & $-0.28\pm 0.02$ \\ \hline
\end{tabular}
\end{center}
\caption{\label{fig:data_crazy} Additional stability diagrams with black dashed lines indicating select pattern shifts. The table contains a summary of shift magnitudes.}
\end{figure}

\newpage
Fig.~\ref{fig:data_crazy} shows an assortment of experimental stability diagrams which have large, but inconsistent, pattern shifts. It is apparent that the characteristics of these shifts vary considerably with the gate potential regime. Some of these diagrams also show evidence of transport comparable to that of double or triple quantum dots~\cite{vanDerWiel}.

\section{Discussion and Conclusion}

One possible explanation for these jumps is that fractional charges are being added to the quantum dot along the shift lines. This would only account for the jumps of a fraction that is favored by a particular theory, for instance $1/3$. Other jumps can then be assigned to trivial charge jumps, and not presented in the paper. However, in these devices, we do not expect fractional charges to manifest because the dot is created in a quasi-one-dimensional semiconductor PbTe nanowire while quantum Hall effects are observed in two-dimensional layers. Furthermore, the dot is measured in the absence of an applied magnetic field, which is typically required for fractionalized particles.

To explain the negative-slope diagonal shifts in Fig.~\ref{fig:data_main} we propose that our device manifests multiple charge traps~\cite{danilov} connected in parallel but still capacitively coupled to the main dot under investigation. Currents through the charge traps is much lower than through the main dot, potentially even zero or one electron per second or a minute. However, the mutual capacitance is serendipitously such that an occupation change in the unintended "dot" leads to a $1/3$-period shift in the primary dot. This is conceptually similar to one of the common approaches for sensing the charge occupation in electron-spin qubits, namely through the control of an additional coupled quantum dot intended solely for charge sensing~\cite{barthel}.

The overall variety of behaviors presented here is in itself another argument against the observation of fractionalized charges, but this evidence could be potentially missed through incomplete analysis. For example, a qualitative analysis of such shifts would likely discard small jumps below 0.2 as being due to trivial effects such as noise, which would also apply to jumps between 0.8 and 1.0. Anything between 0.2 and 0.4, as well as between 0.6 and 0.8, is likely to be perceived as $1/3$. Thus, only jumps that are clearly near 0.5 would be treated as an exception, but these are less frequent: in a purely random distribution these would only occur about 20\% of the time.

We propose that similar concerns affect mesoscopic quantum Hall interferometers. The length scale of these interferometers is often a few hundred nanometers. Multiple gate electrodes, contacts, and dielectric layers may surround each device. A recent study of graphene interferometers like this has found fractional charge jumps in the integer quantum Hall regime~\cite{yang}. 

In analyzing these and similar experiments, it is important to know the landscape of the sample in regards to the possible unintended charge traps in various layers, and how signals from those can coexist with or mimic the exotic signals under investigation. This is possible if sufficient volumes of data are shared at the time of publication, including data extending beyond the regimes depicted in figures from the same sample, data on all samples studied, as well as data taken at different times.

\section{Data availability}

Data are available through Zenodo at DOI: 10.5281/zenodo.8349309.

\section{Duration of study}

The content of this report is based on work done between early 2022 and mid-2023.

\section{Acknowledgments}

The nanowires used in this study are provided by E. Bakkers and S. Schellingerhout from Eindhoven University of Technology. This work is supported by the U.S. Department of Energy Basic Energy Sciences under grant DE-SC-0019274.

\nocite{*}
\bibliography{mainBib}

\end{document}